\documentclass
[aps,prc,nofootinbib,preprint]{revtex4-1}
\usepackage{graphicx,subfigure,latexsym}
\usepackage{amsmath,amssymb,amsfonts}
\usepackage{ulem}
\usepackage{color}
\usepackage{bm}
\usepackage{url,hyperref}

\def \be {\begin{equation} }
\def \ee {\end{equation}}

\def \bem {\begin{multline}}
\def \eem {\end{multline}}

\def \bes {\begin{subequations} }
\def \ees {\end{subequations}}

\def \pd {\partial}

\def \eq {Eq.~}

\def \a {\alpha}

\def \e {\epsilon}
\def \g {\gamma}

\def \o {\omega}

\def \l {\lambda}

\def \s {\sigma}

\def \O {\Omega}

\def \<{\langle}
\def \>{\rangle}
\def \+{\dagger}
\def \({\left(}
\def \){\right)}
\def \[{\left[}
\def \]{\right]}

\def \vp {\bm{p}}

\def \vx {\bm{x}}

\def \vj {\bm{j}}
\def \vA {\bm{A}}

\def \vB {\bm{B}}
\def \vE {\bm{E}}

\def \hp {\hat{\vp}}

\def \hB {\hat{B}_{0}}

\def \pt {p_{\perp}}

\def \EM {\text{EM}}

\def \rel {\text{rel}}

\def \shift {\text{shift}}

\def \phase {(2\pi)^3}

\def \data {\text{data}}

\def \anom {\text{anom}}

\def \cmw {v_{\chi}}

\begin{document}

\title{Electrical conductivity of the quark-gluon plasma and soft photon spectrum in heavy-ion collisions}
\author{Yi~Yin}
\affiliation{
Physics Department,  University of Illinois at Chicago,
Chicago, IL, 60607
}
\affiliation
{
Physics Department, 
Brookhaven National Laboratory, 
Upton, NY 11973
}
\email{yyin@bnl.gov}

\begin{abstract}
We extract the electrical conductivity $\s_0$ of the quark gluon plasma(QGP) and study the effects of magnetic field and chiral anomaly on soft photon azimuthal anisotropy, $v_2$,
based on the thermal photon spectrum at $0.4~GeV<\pt<0.6~GeV$ at RHIC energy.
As a basis for our analysis,
we derive the behavior of retarded photon self energy of a strongly interacting neutral plasma in hydrodynamic regime in the presence of magnetic field and chiral anomaly.
By evolving the resulting soft thermal photon production rate over the realistic hydrodynamic background and comparing the results with the data from the PHENIX Collaboration,
we found that the electrical conductivity at QGP temperature is in the range:
$0.4<\s_0/(e^{2}T) <1.1$,
which is comparable with recent studies on lattice. 
We also compare the contribution from the magnetic field and chiral anomaly to soft thermal photon $v_{2}$ with the data. 
We argue that at LHC,
the chiral magnetic wave would give negative contribution to photon $v_2$.
  \end{abstract}
\maketitle

\section{Introduction}
Photons produced in heavy-ion collisions contain rich information on the properties of quark-gluon plasma(QGP).
The number of photons emitted per unit time per unit volume, 
from a plasma in thermal equilibrium, 
to leading order in $\a_{\EM}$,
is given by\cite{kapusta1993finite}:
\begin{equation}
\label{eq:rate}
\o \frac{d\Gamma_{\g}}{d^{3}\vp} = 
-\frac{1}{\phase}\,
\frac{Im\[\,P^{ij}_T G^{R,\EM}_{ij}(\o, \vp)\,\]}{e^{\o/T} - 1}\big |_{\o= |\vp|}\, . 
\end{equation}
where $P^{ij}_{T}\equiv\delta^{ij}-\hp^{i}\hp^{j}$ is the projection operator.
Here
$G^{ij}_{R,\EM}(\o, \vp)$(i.e. retarded photon self energy) denotes 
the retarded Green's function of the charge current operator $J^{\EM}_{\mu} \equiv \sum_{f}q_f\, \bar{\psi}^{f}\gamma_{\mu}\psi^{f}$.
On the other hand,
the low energy and low momentum behavior of $G^{ij}_{R,\EM}(\o, \vp)$ for any interacting system in the thermal equilibrium is completely fixed by hydrodynamics.
Indeed,
if the conductivity tensor of the system is isotropic, 
i.e. $\s^{ij}=\s_{0}\delta^{ij}$,
the thermal emission rate of soft photons is fully parametrized by $\s_0$:
\begin{equation}
\label{eq:rate-hydro}
\o \frac{d\Gamma_{\g}}{d^{3}\vp} \big |_{\omega=|\vp|}= 
\frac{\a_{\EM}}{\pi^{2}e^{2}}\, \frac{\s_0 \o}{e^{\o/T} - 1}\, . 
\end{equation}

Recently, 
results of low $\pt$ direct thermal photon spectrum at RHIC have been reported by PHENIX Collaboration\cite{Adare:2014fwh,Bannier:2014bja}.
The lowest $\pt$ bin in those results is $0.4~GeV\leq\pt\leq 0.6~GeV$.
It is now well accepted that a near perfect fluid is created in heavy-ion collisions.
The smallness of $\eta/s$ as inferred from relativistic hydrodynamic simulations implies that sQGP enjoys a wider hydrodynamic regime,
to the order of $\pi T$.
Therefore, for photon produced at energy $0.4~GeV\leq\pt\leq 0.6~GeV$,
the hydrodynamic expression,
e.g. \eq\eqref{eq:rate-hydro} does apply.
We then could use \eq\eqref{eq:rate-hydro} to extract $\s_{0}$.
By evolving \eq\eqref{eq:rate-hydro} with the temperature-flow background as generated by solutions of relativistic hydrodynamic equations(cf.~Sec.~\ref{sec:sigma}),
we found at typical QGP temperature(cf.~Fig.~\ref{fig:pt05}): 
\begin{equation}
\label{eq:sigmabound}
0.4 <\frac{\s}{e^{2} T} < 1.1 \, .
\end{equation}
To the extent of our knowledge,
this is the first direct estimation of the electrical conductivity of QGP 
based on soft photon production with realistic hydrodynamics simulation\footnote{
	The conductivity can be related to the diffusive constant via Einstein relation.
The heavy quark diffusive constant in QGP was studied in Ref.~\cite{Moore:2004tg} based on charm spectrum $R_{AA}$ and charm elliptic flow.
Recently,
there is encouraging progress on constraining light quark diffusive constant of QGP 
at cross-over regime by applying fluctuating hydrodynamics in Bjorken expansion to the study of charge density fluctuations in QCD matter\cite{Ling:2013ksb}.
However, we are unaware of any work on directly extracting the conductivity and light quark diffusive constant with the realistic hydrodynamic simulation. 
 }.

On the other hand,
\eq\eqref{eq:rate-hydro} indicates that photon azimuthal anisotropy, $v_2$ should be small in low $\pt$ regime as the effects due to the background elliptic flow are highly suppressed.
However,
the PHENIX results on soft photon $v_2$\cite{Bannier:2014bja} suggest that direct photon $v_2$ 
does not tend to vanish at low $\pt$ limit but saturates at some
positive value(cf.~Fig.~\ref{fig:v2data}).
The non-zero soft photon $v_2$ implies that 
\eq\eqref{eq:rate-hydro} \textit{does} receive sizable modifications in QGP
\footnote{
The corrections due to non-equilibrium would also contribute to soft photon $v_2$.
However, according to the simulation of Ref.~\cite{Shen:2013cca},
the resulting soft photon  $v_2$ is of the order $0.01\sim 0.02$.
}.

One possible source of such modifications is the magnetic field created by the spectator charges of ultra-relativistic heavy ions which can be as large as $eB\sim m_\pi^2$, 
and it points to the perpendicular direction of the reaction plane \cite{Kharzeev:2007jp}.  
In Ref.~\cite{Basar:2012bp}, 
the effects of magnetic field were considered
to explain photon $v_2$ at $\pt>1$~GeV as measured by PHENIX Collaboration\cite{Adare:2011zr}.
In the present paper,
we will study the effects of magnetic field and chiral anomaly on soft photon $v_2$.
As a basis for our analysis,
we will derive the behavior of retarded Green's function $G^{R}_{ij}(\o, \vp)$ in the hydrodynamic regime in a neutral strongly coupled plasma in the presence of homogeneous magnetic field and chiral anomaly.
As triangle anomaly leads to additional terms in the constitute relation of hydrodynamics\cite{Son:2009tf},
the resulting $G^{R}_{ij}(\o, \vp)$ has a much richer structure. 
This opens the possibility to distinguish the effects of chiral anomaly to photon $v_2$.
Furthermore,
as it is not unexpected that the magnetic field will give positive contribution to photon $v_2$,
the phenomenologically important question is how sizable the effects of magnetic field are at heavy-ion collisions.
To answer this question,
a realistic hydrodynamic simulation of photon production is needed.
By evolving the modified soft photon rate in realistic hydrodynamic background,
we found that if the life time of magnetic field $\tau_B>2\sim 3$~fm,
the contribution due to magnetic field to the soft photon $v_2$ is comparable to the experiment results.

This paper is organized as follows. 
Sec.~\ref{sec:GR}
presents the derivation of the behavior of retarded Green's function $G^{ij}_{R}(\o, \vp)$ in the hydrodynamic region in the presence of homogeneous magnetic field and chiral anomaly.
Results are summarized in \eq\eqref{eq:hydrog} and \eq\eqref{eq:Formfactor}.
Though they are the direct consequences of the constitute relation of anomalous hydrodynamics and the linear response theory,
to best of our knowledge,
\eq\eqref{eq:Formfactor} is new in literature .
In Sec.~\ref{sec:sigma}, we extract the electrical conductivity with the realistic hydrodynamic evolution.
Our results are comparable with recent lattice measurement.
In Sec.~\ref{sec:v2B}, 
we investigate the relation between magnetic field and photon $v_2$.
We summarize and conclude in Sec.~\ref{sec:summary}.

\section{Retarded Green's function in the hydrodynamic regime and soft photon production}
\label{sec:GR}

In this section,
we will work out explicitly the behavior of the retarded Green's function $G^{R}_{ij}(\o,\vp)$
of a neutral($n_{V,A}=0$) strongly coupled plasma in the presence of a homogeneous magnetic field$\vB_0$ and chiral anomaly in the hydrodynamic region.
For our purpose,
it is sufficient to consider a plasma with only one flavor with EM charge $q^{f}$.
We start with
the constitute relation for the spatial part of the vector current $J^{\mu}_{V}=q^{-1}_f J^{\mu}_{\EM}$ and axial current $J^{\mu}_{A}$ in a static and homogeneous flow background\cite{Son:2009tf}:
\bes
\label{eq:constituteVA}
\be
J^{l}_{V}= C_{\anom}\mu_{A} B^{l} + 
 q^{-1}_f\sigma^{l m} E_{m}\, -D ^{lm}\nabla_{m}n_{V}\, ,
\ee
\be
J^{l}_{A}= C_{\anom}\mu_{V}B^{l} -D ^{lm}\nabla_{m}n_{A}\, ,
\ee
\ees
where $l,m=1,2,3$ run over spatial components and $\mu_{V},\mu_{A}$ denote chemical potential of vector charge and axial charge respectively.
In \eq\eqref{eq:constituteVA},
the anomaly coefficient
\begin{equation}
C_{\anom} = \frac{q_f N_c}{2\pi^{2}}\, ,
\end{equation}
is defined through the divergence of the axial current
\begin{equation}
\label{eq:anomaly}
\pd_{\mu}j^{\mu}_{A}
=  C_{\anom} q_f \vE\cdot \vB\, .
\end{equation}
$C_{\anom}\mu_{A,V}\vB$ terms in \eq\eqref{eq:constituteVA} are completely induced by chiral anomaly\cite{Son:2009tf}
and are directly related to chiral magnetic effects and charge separation effects\cite{Kharzeev:2007jp,Fukushima:2008xe}(see Ref.~\cite{Kharzeev:2013ffa} for a recent review).
$\sigma^{ij} , D^{ij}$ in \eq\eqref{eq:constituteVA} are conductivity and diffusive tensors in the presence of magnetic field $\vB _{0}$, respectively, and are related by Einstein relation $q^{2}_f\chi D^{ij} =  \sigma ^{ij}$ 
where $\chi$ is the susceptibility.
Due to external magnetic field $\vB_{0}$, 
$\s^{ij}, D^{ij}$, in general, is anisotropic( cf.~\eq\eqref{eq:conductivitytensor}).
 
To determine the behavior of $G^{R}_{ij}(\o, \vp)$ in the hydrodynamic region through the linear response theory,
we now perturb the system by imposing a space-time dependent vector potential
$\delta\vA \propto e^{- i \o t + i \vp\cdot\vx}$.
Due to \eq\eqref{eq:constituteVA} and $\vE=-\pd_{t}\vA$,
for a neutral plasma,
the change of current $\delta \vj_{V,A}$ in response to $\delta\vA$ now reads:
\begin{equation}
\label{eq:deltaj}
\(
\begin{array}{c}
  \delta j^{l}_V(\o, \vp)\\
  \delta j^{l}_A(\o, \vp)
\end{array}
\) 
=
(- i p_{m})
\(
\begin{array}{cc}
 D^{m l} ,\,& D^{m l}_{5}\\
 D^{m l}_{5},\,& D^{ml}
\end{array}
\)
\(
\begin{array}{c}
 \delta n_{V}(\o, \vp)\\
 \delta n_{A}(\o, \vp)
\end{array}
\)
+ i \o \(
\begin{array}{c}
\s^{lm}\\
  0
\end{array}
\)  q^{-1}_f\delta A_{m}(\o, \vp)
\, .
\end{equation}
Here we relate $\delta \mu_{V,A}, \delta n_{V,A}$ by
$\chi \delta \mu_{V,A} =  \delta n_{V,A}$ where $\chi$ is the susceptibility.
We expect that in the chirally symmetric phase,
the susceptibility for the axial charge and the vector charge are approximately identical.
For future convenience,
we have also introduced:
\begin{equation}
D^{ij}_{5} \equiv \frac{i\cmw}{p}\hp^{i}\hB^{j}\, ,
\qquad
\s^{ij}_{5} \equiv q^{2}_f\chi D^{ij}_{5}
=\frac{i C_{\anom}B_{0}}{p}\hp^{i}\hB^{j}\, . 
\end{equation}
where the speed of chiral magnetic wave\cite{Kharzeev:2010gd} reads:
\begin{equation}
\label{eq:cmw}
\cmw \equiv \frac{C_{\anom} B_0}{\chi}\, . 
\end{equation}

From anomaly equation \eq\eqref{eq:anomaly} and conservation of vector charge $\pd_{\mu}j^{\mu}_{V}=0$,
we also have
\begin{equation}
\label{eq:chargeVA}
\(
\begin{array}{cc}
\o+i p_{l}p_{m}D^{m l} ,\,& i p_{l}p_{m} D^{m l}_{5}\\
i p_{l}p_{m} D^{m l}_{5},\,& \o+i p_{l}p_{m}D^{m l}
\end{array}
\)
\(
\begin{array}{c}
  \delta n_V(\o, \vp)\\
  \delta n_A(\o, \vp)
\end{array}
\) 
=
i\o p_{l}
\(
\begin{array}{c}
\s^{lm}\\
 \s^{lm}_{5}
\end{array}
\)
 q^{-1}_f\delta A_{m}
\, .
 \end{equation}
Now solving for $\delta n_{V,A}$ in terms of $\delta\vA$ in \eq\eqref{eq:chargeVA}
and put them back in the expression of $\vj_{V}$ in \eq\eqref{eq:deltaj},
one arrives at:
\begin{equation}
\delta j^{l}_{V}
= 
i\o\[\, 
   \(p_{r}D^{rl},p_{r}D^{rl}_{5}\)\(
\begin{array}{cc}
i\o- p_{i}p_{j}D^{i j} ,\,& -p_{i}p_{j} D^{i j}_{5}\\
-p_{i}p_{j} D^{ij }_{5},\,& i\o- p_{i}p_{j}D^{ij}
\end{array}
\)^{-1}
\(
\begin{array}{c}
 p_{q}\s^{qm}\\
p_{q}\s^{qm}_{5}
\end{array}
\)
+ \s^{lm}\] q^{-1}_{f}\delta A_{m}\, .
\end{equation}

The tensor $\sigma^{ij}$ in the presence of $\vB_{0}$ may be decomposed as\cite{lifshitz1981physical}:
\begin{equation}
\label{eq:conductivitytensor}
\sigma^{ij} = \s_{0} \delta^{ij} 
-\s_{B,T} ( \delta^{ij}-\hB ^{i}\hB ^{j}) +
\s_{B,L} \hB ^{i}\hB ^{j} 
=q^{2}_f\chi \[\,D_{0} \delta^{ij} 
-D_{B,T} ( \delta^{ij}-\hB ^{i}\hB ^{j}) +
D_{B,L} \hB ^{i}\hB ^{j} \, \] \, ,
\end{equation}
where $\hB$ is the directional vector of $\vB_0$.
Here $\s_0$ denotes the conductivity in the absence of magnetic field,
$\s_{B,L},\s_{B,T}$ denote the change of conductivity in the longitudinal and transverse direction of magnetic field $\vB_0$ 
\footnote{
If the Hall conductivity $\s_{H}$ is non-zero, 
one could add an additional term $i \s_H \e_{ijk} \hat{B}^{k}_{0}$ to $\s_{ij}$.
}.
According to linear response theory:
\begin{equation}
\< J^{\mu}_{V}(\o, \vp)\>_{\vA}
= - G^{\mu\nu}_{R,\EM}(\o, \vp)\, (q^{-1}_fA_{\nu})\, ,
\end{equation}
we then obtain:
\begin{equation}
\label{eq:hydrog}
G^{ij}_{R,\EM}(\o , \vp )
=\, F_{L}(\o , \vp )\,\hp^{i}\hp^{j}+ 
F_{T}\,(\o , \vp)P^{ij}_{T} +
	F_{pB}\,(\o , \vp )\(\,\hp^{i}\hB^{j}+\hp^{j}\hB^{i}\,\) 
	+ F_{BB}\,(\o , \vp )\hB^{i}\hB^{j}\,,
\end{equation}
where form factors in \eq\eqref{eq:hydrog} are given by:
\bes
\label{eq:Formfactor}
\begin{equation}
F_{T} = -i\o \(\s_{0}-\s_{B,T}\)\, ,
\end{equation}
\begin{equation}
F_{L}=
 \frac{-i\o\(\s_{0}-\s_{B,T}\)}{2}
\{
\frac{ \o -\cmw p (\hp\cdot\hB)+ i p^{2}(D_{B,T}+D_{B,L})(\hp\cdot\hB)^{2}}
{\O(\o,\vp,\hB)}
+\(\hB\to -\hB\)\}\, ,
\end{equation}
\begin{equation}
F_{pB} =
 \frac{-i\o\(\s_{0}-\s_{B,T}\)}{2}
\{\,
\frac{
 \cmw p - i p^{2}(D_{B,T}+D_{B,L})(\hp\cdot\hB)^{2}}
{\O(\o,\vp,\hB)}
+\(\hB\to -\hB\)\,\}\, ,
\end{equation}
\begin{equation}
F_{BB} = 
-\frac{i\o}{2}\{\,
\frac{
 i\cmw^{2}\chi+ 
 (\s_{B,T}+\s_{B,L})\[\o+\cmw p (\hp\cdot\hB)+i p^{2}(D_{0}-D_{B,T})\] 
  }
{\O(\o,\vp,\hB)}
+\(\hB\to -\hB\)\,\}\, ,
\end{equation}
\ees
where
\begin{equation}
\O(\o,\vp,\hB)
=\o - \cmw p (\hp\cdot\hB)+
i p^{2}\[(D_{0}-D_{B,T})+(D_{B,T}+D_{B,L})(\hp\cdot\hB)^{2}\]\, .
\end{equation}
In \eq\eqref{eq:Formfactor}, 
the contributions due to right-handed chiral fermions have been written down explicitly in the brackets $\{\ldots\}$ while those due to left-handed chiral fermions are easily obtained by replacing $\hB$ with $-\hB$ as denoted by $(\hB\to -\hB)$ in \eq\eqref{eq:Formfactor}.

As one can check,
in the absence of $\vB_{0}$ that $\s_{B,T},\s_{B,L},\cmw=0$,
$F_{BB},F_{pB}$ vanish and we recover the well-known results:
\begin{equation}
\lim_{\vB_{0}\to 0}F_T(\o,p)= -i\o \sigma_0\, 
\qquad
\lim_{\vB_{0}\to 0}F_L(\o,p)= \frac{-i\o^{2} \sigma_0}{\o + i D_0 p^{2}}\, .
\end{equation}

Returning to \eq\eqref{eq:hydrog} and \eq\eqref{eq:Formfactor},
we see immediately that $G^{\text{R}}_{ij}(\o , \vp)$ has poles when $\O(\o,\vp,\pm\hB)=0$.
The corresponding dispersion relation is:
\begin{equation}
 \label{eq:cmwdispersion}
 \o(\vp) = \pm \cmw p (\hB \cdot \hp)
 -ip^2
  \[\, \(D_0 -D_{B,T} \) + \(D_{B,L}+D_{B,T}\) (\hB \cdot \hp)^2\,\] .
\end{equation} 
For $\hB \cdot \hp = 0$,
\eq\eqref{eq:cmwdispersion}
describes the conventional diffusive modes
while for $\hB \cdot \hp \neq 0$, \eq\eqref{eq:cmwdispersion}
describes a propagating hydrodynamical mode, namely, chiral magnetic wave\cite{Kharzeev:2010gd}.
We point out here that due to chiral magnetic wave poles, 
zero frequency limit and zero momentum limit of $G^{ij}_{R,\EM}(\o,\vp)$ may not commute with each other.
Special care may be needed when apply Kubo formula to a plasma in the presence of magnetic field and chiral anomaly. 

To determine photon production rate in the hydrodynamic region,
we only need to know the imaginary part of $F_{T}(\o, \vp), F_{BB}(\o,\vp)$ along the light-cone $\o=|\vp|$:
\begin{equation}
\label{eq:ImpG}
Im(P^{ij}_{T}G^{R}_{ij}(\o,\vp))\big |_{\o=|\vp|}
= 2 Im \[ F_{T}(\o = |\vp|)\] + Im \[F_{BB}(\o=|\vp|)\]
\[\,1-(\hp\cdot\hB)^{2}\,\]\, . 
\end{equation}
Keeping terms of the lowest order in $\o$ in \eq\eqref{eq:ImpG},
we then have, in the presence of magnetic field and chiral anomaly, that:
\begin{multline}
  \label{eq:rateB}
Im(P^{ij}_{T}G^{R}_{ij}(\o,\vp))\big |_{\o=|\vp|}
= \frac{\a_{\EM}\,\o }{2\pi^{2}e^{2}}\,
\{
2 \(\s_{0}-\sigma_{B,T}\)
+ \\
\(\, 1-(\hp\cdot\hB)^{2}\,\)
\,\[\, 
       \(\s_{B,T}+\sigma_{B,L}\)
 +\cmw ^2
        \(\s_{0}-\sigma_{B,T}\)\,
    \] \,
\frac{1+\cmw^{2}(\hp\cdot\hB)^{2}}
{\(1-\cmw^{2}(\hp\cdot\hB)^{2}\)^{2}} \, 
\}
\end{multline}
and in the absence of $B$,
we recover \eq\eqref{eq:rate-hydro}.

In the following sections,
we will use \eq\eqref{eq:rate-hydro} and \eq\eqref{eq:rateB} to study soft photon production in heavy-ion collisions.

\section{Electrical conductivity of QGP}
\label{sec:sigma}
The thermal photon momentum spectrum produced during the evolution of the radiating fireball can be written as:
\begin{equation}
\o\frac{dN_{\gamma}}{d^3\vp}
= \int d^4x \, \o_{\shift}\frac{d\Gamma_{\g}}{d^{3}\vp '}\big|_{\o_{\shift}=|\vp'|}\, 
\end{equation}
where the photon energy, 
which is $\o$ in the lab frame,
is red-shifted to $\o_{\shift}$ in the frame that fluid is at rest:
\begin{equation}
\o_{\shift} = p^{\mu}u_{\mu} = 
\o u^{\tau}\[\cosh(\xi-Y) - v_x\cos\phi_p -v_y\sin\phi_p \]\, .
\end{equation}
Here, 
assuming the boost-invariance,
the 4-velocity of the flow field is
$u^{\mu}=(u^{t},u^{z},u^{x},u^{y}) = u^{\tau}\(\cosh\xi,\sinh\xi, v_{x},v_y\,\)$.
We use Bjorken's coordinates $\tau,\xi, x, y$, with 
$\tau =\sqrt{t^{2}-z^{2}}$ the longitudinal proper time 
and $\xi =\tanh^{-1}(z/t)$ the space-time rapidity that $d^{4}x = \tau d\tau d\xi dx dy $.
The photon momentum is parametrized by its rapidity $Y$, 
transverse momentum $\pt$ and  azimuthal  emission angle $\phi_{p}$, 
i.e., 
$p^{\mu} = \pt \(\cosh Y,\sinh Y, \cos\phi_{p},\sin\phi_{p}\)$.

In the present section,
we will estimate the value of the electrical conductivity by neglecting possible modifications due to the magnetic field.
We will return to the effects of magnetic field in the next section.
For the soft photon production at heavy-ion collisions,
we have from \eq\eqref{eq:rate-hydro} that:
\begin{equation}
\label{eq:low-pt-rate}
\frac{dN_{\gamma}}{d^{2}\pt dY} 
= \,\frac{\a_{\EM}}{\pi^{2}}\, 
\int d^{4}x\, \(\frac{\s_0}{e^{2}T}\)
\, \frac{\o_{\shift}\, T }{\exp(\frac{\o_{\shift}}{T})-1}\, .
\end{equation}
We will concentrate on the photo productions at mid-rapidity $Y=0$ and expand the photon production in Fourier Harmonics:
\begin{equation}
\frac{dN_{\gamma}}{d^{2}\pt dY} =
\frac{dN_{\gamma}}{2\pi \pt d\pt dY}\, 
\[\,1 + 2 v_2(\pt)\cos(2\phi_p)+\ldots\,\]\, .
\end{equation}
We therefore have:
\begin{equation}
\frac{dN_{\gamma}}{2\pi \pt d\pt dY}
= \frac{\a_{\EM}}{\pi^{2}}\,\int^{2\pi}_{0} \frac{d\phi_{p}}{2\pi} 
\int d^{4}x\, \(\frac{\s}{e^{2}T}\)
\, \frac{\o_{\shift}\, T }{\exp(\frac{\o_{\shift}}{T})-1}\, .
\end{equation}
Now introducing the dimensionless quantity:
\begin{equation}
\label{eq:sigmaaverage}
\<\frac{\s}{e^2T}\>_{\text{QGP}} \equiv 
\frac{
  \frac{\a_{\EM}}{\pi^{2}}\,
    \int^{2\pi}_{0} \frac{d\phi_{p}}{2\pi}  
 \int d^{4}x\, \(\frac{\s}{e^{2}T}\)
\, \frac{\o_{\shift}\, T }{\exp(\frac{\o_{\shift}}{T})-1}
    }
    {\frac{\a_{\EM}}{\pi^{2}}\, 
\int^{2\pi}_{0} \frac{d\phi_{p}}{2\pi} \int d^{4}x 
\, \frac{\o_{\shift}\, T }{\exp(\frac{\o_{\shift}}{T})-1}
     }
= \frac{\(\frac{dN_{\gamma}}{2\pi\pt d\pt dY}\)_{\data}}
 {\frac{\a_{\EM}}{\pi^{2}}\, 
\int^{2\pi}_{0} \frac{d\phi_{p}}{2\pi}\int  d^{4}x\, 
\, \frac{\o_{\shift}\, T }{\exp(\frac{\o_{\shift}}{T})-1}
     }\, .
\end{equation}
As the conductivity in hadronic phase is much smaller than that in QGP state due to the reduction of the charge carriers in the medium,
$\<\s/e^2T\>_{\text{QGP}}$ provides us an estimation of $\s/e^{2}T$ at typical QGP temperature. 

\begin{figure}[htb]
 \centering
\includegraphics[width=.6\textwidth]{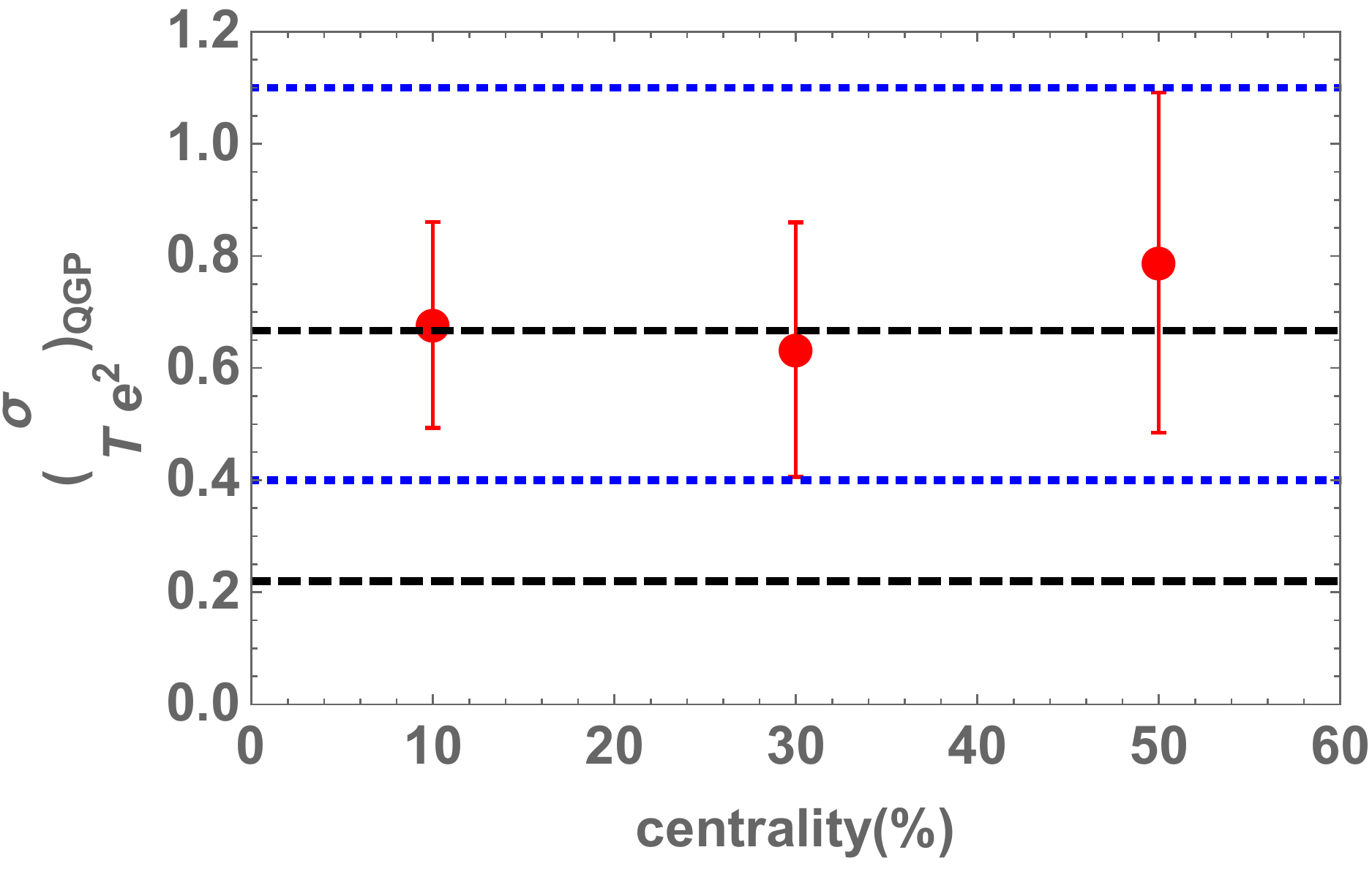}
\caption{
 \label{fig:pt05}
(Color Online)
$\<\s/eT^2\>_{QGP}$(red dots), 
the average of $\s/e^{2}T$ in QGP as computed from \eq\eqref{eq:sigmaaverage} using photon production at $\pt$ bin $0.4~GeV <\pt<0.6 GeV$ 
for three different centrality bins
($0 - 20\%$, $20-40\%$, $40-60\%$)\cite{Adare:2014fwh}.
Blue dotted lines are corresponding to the upper and lower bound for
$\s/e^2T$ as quoted in the abstract and in \eq\eqref{eq:sigmabound}. 
Dashed horizontal lines are corresponding to the range of $\s/(e^{2}T)$ as estimated by lattice simulation in Ref.~\cite{Ding:2010ga} with $C_{\EM}=(2/3)e^{2}$.
}
\end{figure}

In Fig.~\ref{fig:pt05},
we show $\<\s/e^2T\>_{\text{QGP}}$,
the average $\s/(e^{2}T)$ in QGP as defined by \eq\eqref{eq:sigmaaverage}.
The direct photon production data(after subtraction of hard-scattering component)
are taken from results by the PHENIX Collaboration\cite{Adare:2014fwh} 
at $\pt$ bin $0.4~\rm{GeV}<\pt<0.6~\rm{GeV}$ in $\sqrt{s_{NN}}=200$~GeV $Au+Au$ collisions.
%\footnote{
%One may worry about the applicability of 
%\eq\eqref{eq:rate-hydro} to that $\pt$ bin at the late stage of the fireball expansion due to the dr%opping of temperature.
%However, the shifted frequency $\o_{\shift}$ is also smaller due to
%the strong radial flow at that stage. }
The denominator of the last term of \eq\eqref{eq:sigmaaverage} is evaluated at $\pt =0.5$~GeV using realistic hydrodynamic background. 
To model such background,
we employ results computed with ``VISH2+1", 
a viscous hydro code,
developed by Huichao Song and U. Heinz\cite{Song:2007fn,*Song:2007ux,*Song:2008si},
in $2+1$ dimensions assuming longitudinal boost invariance.
Those simulations,
which reproduce hadron spectrum in the experiment well, 
were performed by Chun Shen\cite{Shen:2010uy,*Renk:2010qx}
and the results are accessible to the public via the website:
https://wiki.bnl.gov/TECHQM/index.php.
The hydrodynamic evolution starts at $\tau_i=0.4$~fm and ends on an isothermal surface at $T_{\rm{dec}}=130$~MeV with $\eta/s=0.20$ and the lattice-based equation of state ``s95p-PCE" \cite{Huovinen:2009yb,Shen:2010uy}.

We have performed our analysis for three different centrality bins: $0 - 20\%$, $20-40\%$, $40-60\%$
\footnote{
Other parameters to generate background hydrodynamic flow include the initial entropy density $s_{i}=86.7~\rm{fm}^{-3}$ for all impact parameters.
Results shown in the current paper are using Glauber initial conditions. 
We have performed the calculation for both CGC initial conditions and Glauber initial conditions at $b=3.16$~fm and found a minor difference. 
We take impact parameters $b=3.16, 5.78, 7.49,.8.87,10.1,11.1$~fm which correspond to centrality ranges $0-10\%, 10-20\%,20-30\%,30-40\%,40-50\%,50-60\%$ respectively.
}
.
As the conductivity $\s_0$ reflects the transport properties of QGP and the effective temperature as extracted from thermal photon spectrum for those three centrality bins are similar\cite{Adare:2014fwh},
we expect that $\<\s/(e^2T)\>_{\text{QGP}}$ would have a weak dependence on centrality. 
This is indeed the case as one can see in Fig.~\ref{fig:pt05}:
while both soft photon production and hydrodynamic backgrounds are different for those centrality bins,
the resulting $\<\s/(e^2T)\>_{\text{QGP}}$ shows little dependence on the centralities.    

The error bars shown in Fig.~\ref{fig:pt05} are determined from the
experimental (systematic) uncertainties in the photon production as $\<\s/(e^2T)\>_{\text{QGP}}$ given in \eq\eqref{eq:sigmaaverage} linearly depends on photon production measured in experiment.
On the theory side,
the major source of uncertainty is from the correction to \eq\eqref{eq:rate-hydro} at $\pt=0.5$~GeV.
One may get an idea on the magnitude of such corrections from strongly coupled QCD-like theories. 
For example,
for ${\cal N}=4$ super Yang-Mills theory in the strongly coupling limit, 
the corrections to \eq\eqref{eq:rate-hydro} is at most $20\%$ for $\pt$ from $0$ to $\pi T$(see Fig.~1 of Ref.~\cite{CaronHuot:2006te}).
In our calculations,
we did not include the contributions at pre-equilibrium stage. 
To estimate the resulting uncertainty,
we have extrapolated from the 
initial time $\tau_{i}=0.4$~fm to a $3$ times smaller value
assuming 1-dimensional boost-invariant expansion between these times
and computed the photon production during that interval.
The corrections is a few percent at most.

We now compare our results with the electrical conductivity as measured on lattice. 
A recent quenched study using Wilson-Clover fermions\cite{Ding:2010ga}
in the continuum limit found that: 
 $0.33 C_{\EM} \leq \s/T \leq 1 C_{\EM} $ at $T=1.45T_c$.
This result is consistent with other lattice measurements\cite{Brandt:2012jc,*Amato:2013naa,*Aarts:2007wj}.
Here $C_{\EM}=\sum_{f}q^{2}_f$ counts number of charge carriers.
For example, for $f=u,d$, $C_{\EM} = (5/9) e^{2}$ and for $f=u,d,s$, $C_{\EM}=(2/3) e^{2}$.
For comparison, we plot the range of $\s/(e^{2}T)$ at $T=1.45T_c$ as indicated in Ref.~\cite{Ding:2010ga} in Fig.~\ref{fig:pt05} in dashes horizontal lines with $C_{\EM}=(2/3)e^{2}$ by assuming in QGP, $u,d,s$ all contribute to the conductivity. 
It is seen there that our results are completely comparable with
lattice measurement.
Our results is also consistent with
Ref.~\cite{Cassing:2013iz} using the off-shell parton-hadron-string
dynamics transport approach. 

\section{Soft photon $v_2$, chiral anomaly and magnetic field}
\label{sec:v2B}

In this section,
we will study the effects of magnetic field and chiral anomaly on soft photon $v_2$. 
We will evolve soft photon production rate in the presence of magnetic field and chiral anomaly \eq\eqref{eq:rateB} as derived in Sec.~\ref{sec:GR} in the realistic hydrodynamic background.
As at RHIC energy,
the typical speed of chiral magnetic wave $\cmw$ is around $0.1\sim 0.4$\cite{Yee:2013cya} using the susceptibility measured on lattice\cite{Borsanyi:2011sw},
we then neglect $\cmw^{2}$ terms in \eq\eqref{eq:rateB} and approximate the photon rate at mid-rapidity $Y=0$ as:
\begin{multline}
  \label{eq:rateB1}
(\frac{\o}{e^{\o/T}-1})^{-1}\[\o \frac{d\Gamma_{\g,B}}{d^{3}\vp}\]
\approx \frac{\a_{\EM}}{2\pi^{2}e^{2}}\,\[ \,
2 \(\s_{0}-\sigma_{B,T}\)
+\(\s_{B,T}+\sigma_{B,L}\) \cos ^{2}\phi_{\vp}
\,
\] + {\cal O}(\cmw^{2})\\
=\frac{\a_{\EM}\s_0}{\pi^{2}e^{2}}\,\[ \,
(1 - \frac{3}{4}r_T + \frac{1}{4}r_L) + 
\frac{1}{4}(r_T + r_L)\cos(2\phi_{\vp})
\]\, .
\end{multline}
Here, we have introduced dimensionless ratio $r_T, r_L$:
\begin{equation}
r_T \equiv \s_{B,T}/\s_{0} \, ,
\qquad
r_L \equiv \s_{B,L}/\s_{0} \, 
\end{equation}
to characterize the relative change of conductivity in the presence of magnetic field.

We now estimate the contribution from the magnetic field to photon $v_2$ as:
\begin{equation}
\label{eq:v2B}
v_2(B) \approx
\frac{\frac{\a_{\EM}}{8\pi^{2}}\, 
\int^{2\pi}_{0} \frac{d\phi_{p}}{2\pi}\int  d^{4}x\, 
\, \frac{\o_{\shift}\, T }{\exp(\frac{\o_{\shift}}{T})-1}(r_T + r_L )
     }
     {\frac{\a_{\EM}}{\pi^{2}}\, 
\int^{2\pi}_{0} \frac{d\phi_{p}}{2\pi}\int  d^{4}x\, 
\, \frac{\o_{\shift}\, T }{\exp(\frac{\o_{\shift}}{T})-1}(1 - \frac{3}{4}r_T + \frac{1}{4}r_L )
     } \, . 
\end{equation}

To compute $v_{2}(B)$, 
we need to determine $r_T$ and $r_L$.
Let us first consider $r_T=\s_{B,T}/\s_{0}$ under Drude approximation(relaxation time approximate).
Recall the equation of motion for a massive particle in the presence of EM field and a drag force: 
\begin{equation}
\frac{d\vp}{d t} = q_{f} \vE + \frac{\vp}{M}\times q_f\vB - \frac{\vp}{\tau_{\rel}}\, ,
\end{equation}
where $\tau_{\rel}$ denotes the relaxation time.
By imposing the steady-state condition $\frac{d\vp}{d t}=0$ and computing the current in response to $\vE$,
one finds:
\begin{equation}
\label{eq:rTkinetic}
r_T= \frac{\s_{B,T}}{\s_{0}}
= \frac{\(q_f B \tau_{\rel}/M\)^{2}}
{1+\(q_f B \tau_{\rel}/M\)^{2} }\, ,
\qquad
r_L= \frac{\s_{B,T}}{\s_0}=0\, . 
\end{equation} 
As charge carriers moving along the direction of magnetic field $\vB$ do not feel the Lorentz force,
the magnetic field would not affect the longitudinal components of
conductivity tensor, i.e. $r_L=0$ under the drude estimation.
Chiral anomaly may introduce an non-trivial contribution to the
longitudinal conductivity. 
However, as the purpose of this section is to estimate the effects of
magnetic field to soft photo $v_2$, 
we will defer the effects due to $r_L$ to future studies.

We now ready to evaluate $v_{2}(B)$ as defined by \eq\eqref{eq:v2B} in the realistic hydrodynamic background as we did in the previous section.
To estimate $r_T$ using \eq\eqref{eq:rTkinetic},
we need to estimate $\tau_{\rel}/M$ in QGP.
In ${\cal N}=4$ supersymmetric Yang-Mills (sYM) theory in strong
coupling limit, 
this is known for heavy quarks\cite{Herzog:2006gh,*CasalderreySolana:2006rq,*Gubser:2006bz},
\be
\label{eq:sYMtau}
\(\frac{\tau_{\rel}}{M} \)_{sYM}= 
\frac{2}{\sqrt{g^2N_c} \pi T^2}\, . 
\ee
Following \cite{Gursoy:2014aka},
we will use \eq\eqref{eq:sYMtau} with $g^2N_c=6\pi$ .
We will parametrize our ignorance of $\tau_{\rel}/M$ in QGP by
introducing a dimensionless parameter $\l$:
\begin{equation}
\label{eq:oB}
\(\frac{\tau_{\rel}}{M} \)_{QGP}
=
\l \(\frac{\tau_{\rel}}{M} \)_{sYM}
\, . 
\end{equation}
We will treat $\l$ as a free parameter and study the effects of the magnetic field with various $\l$s.
As  $r_T$ depends on the $q_{f}$. 
Strictly speaking, 
in \eq\eqref{eq:v2B},
one should sum the contributions from different flavors.
However, as the photon rate is proportional to $q^{2}_f$,
the number of photon produced by $u$ quarks is roughly four times that produced by $d$ quarks.
We therefore, in our actual evaluation of \eq\eqref{eq:v2B},
set $q_f=q_u=(2/3)e$.

\begin{figure}
	\includegraphics[width=.90\textwidth]{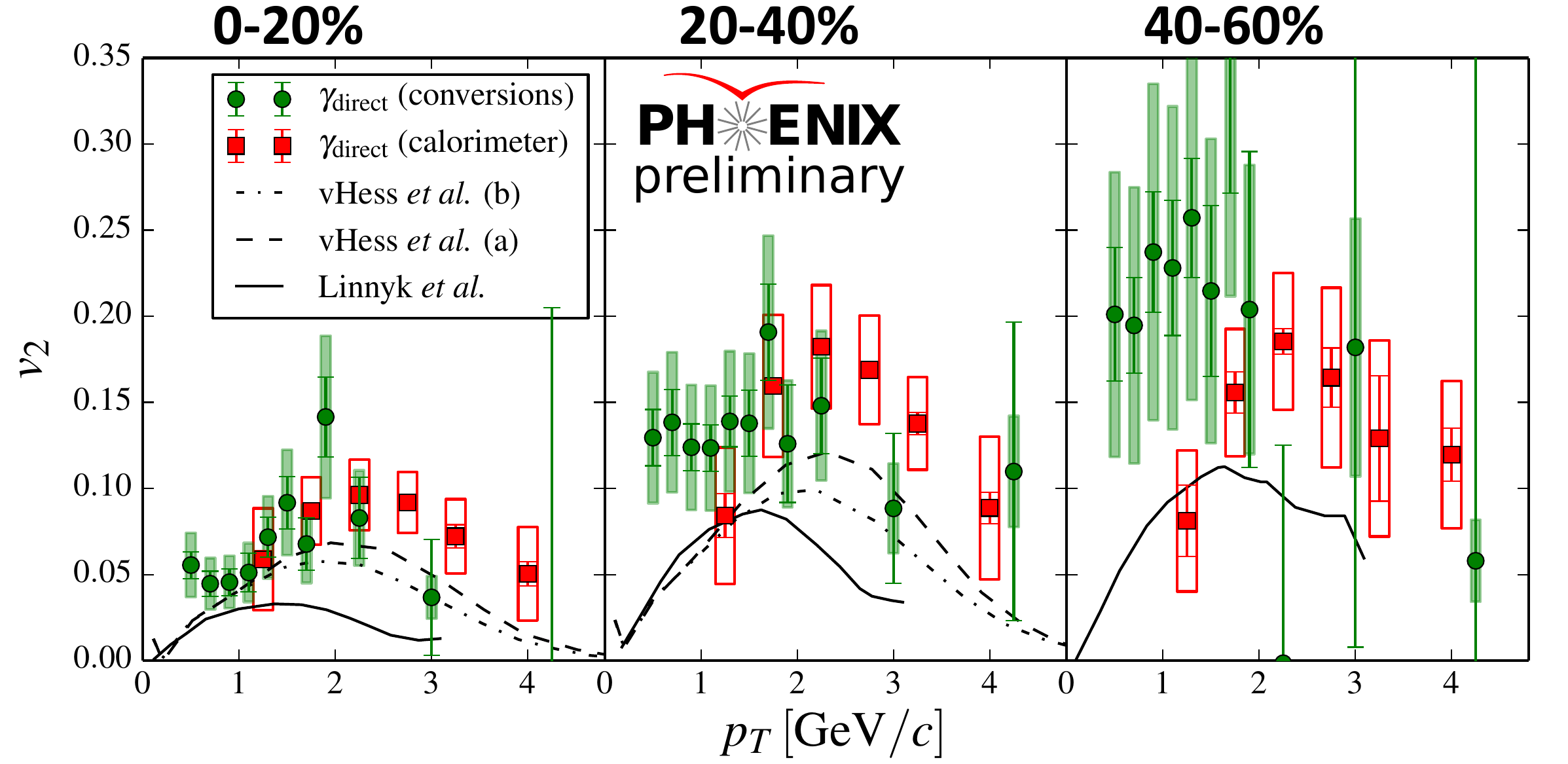}
\caption{
  \label{fig:v2data}
(Color Online)
PHENIX results of direct photon $v_2$
\cite{Bannier:2014bja} (Figure is reproduced from Ref.~\cite{talk2014}). 
}
\end{figure}
We finally specify the profile $eB$ during the hydrodynamic evolution.
We  neglect  the  spatial  gradients  of  magnetic  field  and
take it in the lab frame along the $y$ direction.
We use the time-varying profile of the magnetic field with a parametrization
\be
eB(\tau)=\frac{(eB)_{\rm max}}{1+(\tau/\tau_B)^2}\, ,
\ee 
where we call $\tau_B$ the lifetime of the magnetic field. 
This form has been used in previous literature widely (see, for example, Ref.~\cite{Basar:2012bp,Yee:2013cya}).
We take $(eB)_{\rm max}= 3,5,7,8,9,10 m^{2}_{\pi}$ for
$b=3.16,5.78,7.49,8.87,10.1,11.1~\rm{fm}$
as guided by Ref.~\cite{Bzdak:2011yy}. 
Due to current controversy over the medium effects on $\tau_B$\cite{McLerran:2013hla,Tuchin:2013apa},
we will leave $\tau_B$ as a free parameter in the following calculations
\footnote
{
It should also be pointed out that if $\tau_B<<\tau_{\rel}$,
the hydrodynamic expression \eq\eqref{eq:rateB1} does not apply.
}
. 

\begin{figure}
	\subfigure{
			\label{fig:v20020}
	\includegraphics[width=.30\textwidth]{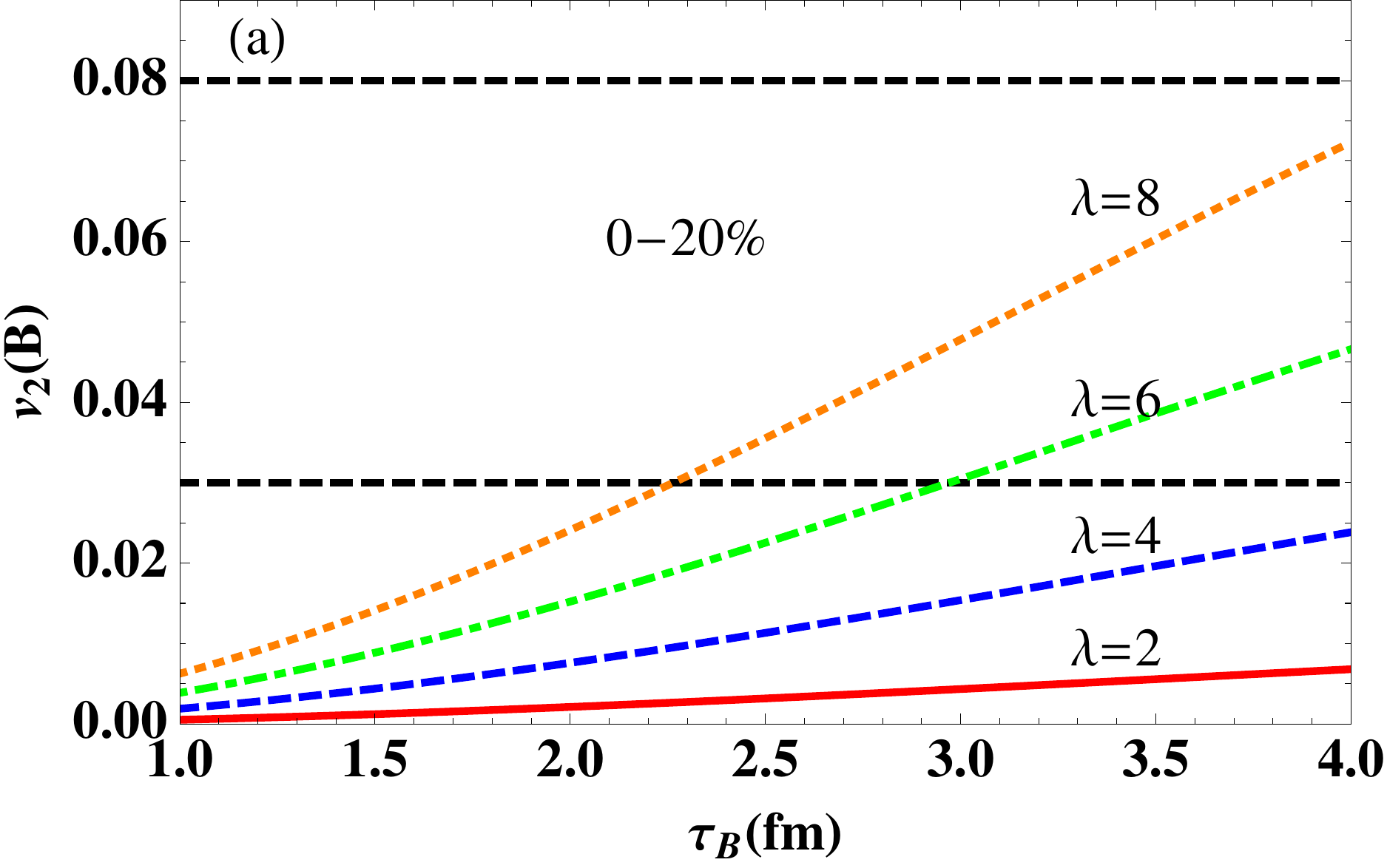}
	    }
	\subfigure
	    {
	      \label{fig:v2t2040}
				\includegraphics[width=.30\textwidth]{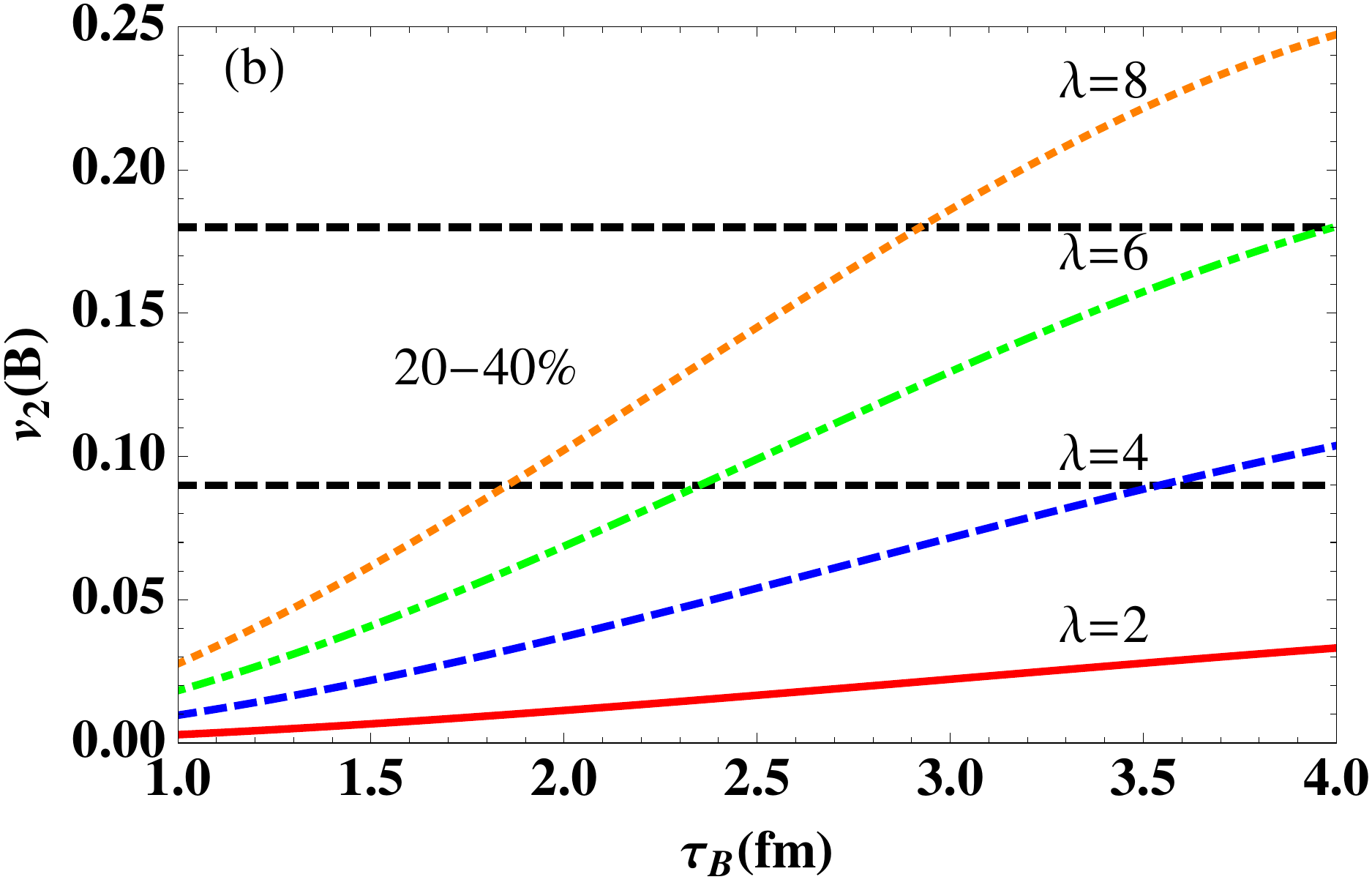}
		}
                	\subfigure
	    {
	      \label{fig:v24060}
				\includegraphics[width=.30\textwidth]{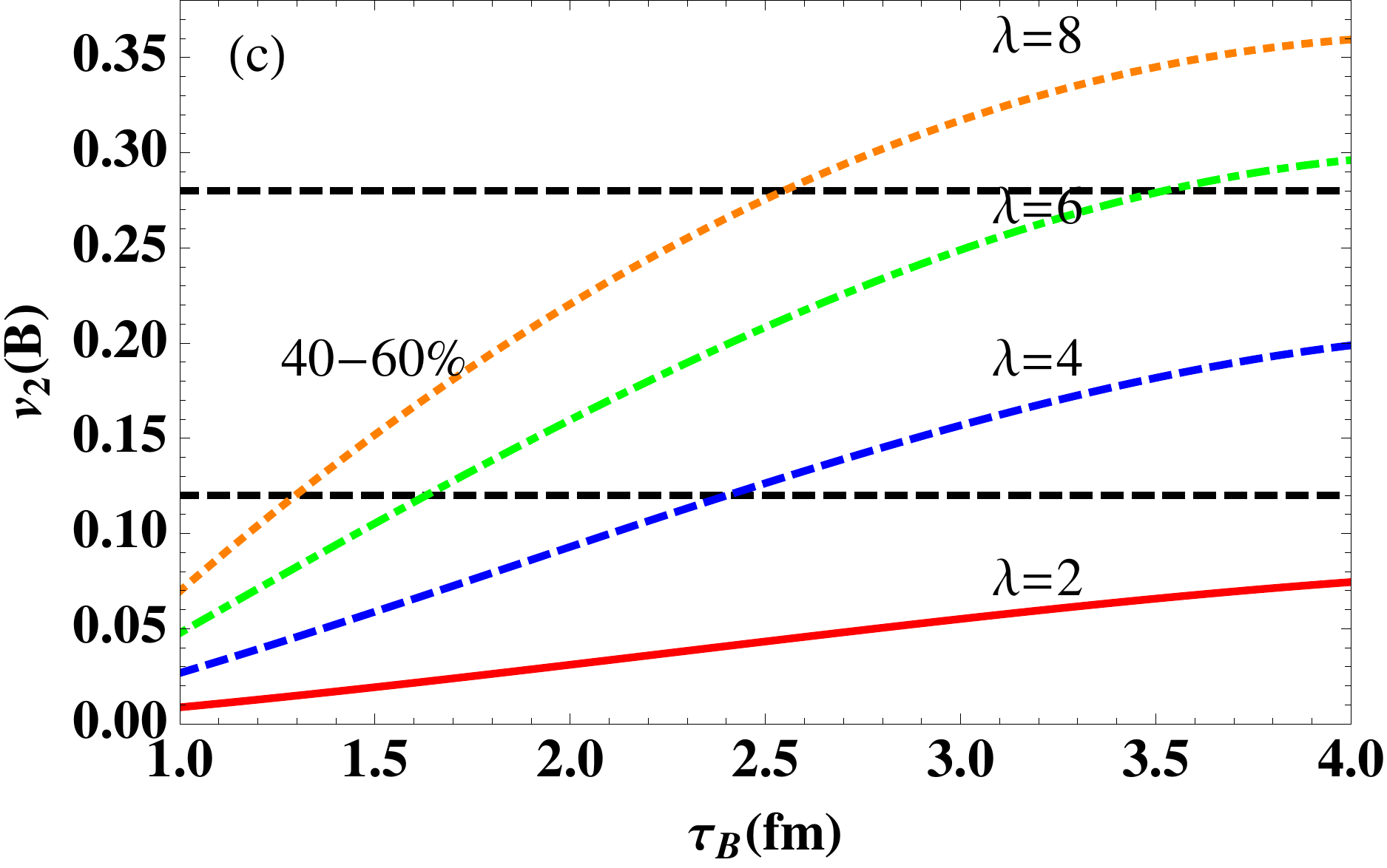}
		}
\caption{
  \label{fig:v2hydro}
(Color Online)The contribution of magnetic field to photon
$v_2$,$v_2(B)$(cf.~\eq\eqref{eq:v2B}), at $\pt=0.5$~GeV vs the life time of magnetic field $\tau_{B}$.
Photon production is  computed based on \eq\eqref{eq:rateB} and \eq\eqref{eq:rateB1} for four different $\l=2,4,6,8$ (red solid, blue dashed, green dotdashed, orange dotted curves respectively).
Here $\l $ is a parameter appearing in \eq\eqref{eq:oB} which parametrizes the $\o_{B}\tau_{\rel}$ in the plasma. 
Two dashed horizontal lines correspond to the upper and lower bound of
direct photon $v_2$ at $\pt=0.5$~GeV from the data(c.f.~Fig.~\ref{fig:v2data}).
Subfigure (a),(b),(c) are corresponding to centrality bins $0-20\%,20-40\%,40-60\%$ respectively.
}
\end{figure}
We have computed photon $v_2(B)$,
the contribution from magnetic field to photon $v_2$,
as a function of the life time of magnetic field $\tau_B$,
by evolving the soft photon production rate in the presence of magnetic field based on hydrodynamics, 
i.e. \eq\eqref{eq:rateB} and \eq\eqref{eq:rateB1}.
$r_T$ appearing in \eq\eqref{eq:rateB1} is taken from \eq\eqref{eq:rTkinetic} and $\o_B\tau_{\rel}$ is given by \eq\eqref{eq:oB}.
We have present our results for three different centrality bins
($0-20\%,20-40\%,40-60\%$) with four different $\l=2,4,6,8$.
The dependence of $v_2(B)$ on $\tau_B$ and $\l$ are similar for the
centrality bins under study.
Perhaps not surprising,
the contribution of magnetic field to photon $v_2$ increases with growing $\l$.
\footnote{
As $\l$ here is the ratio of the actual $\phi_{B}$ in plasma to the characteristic $\phi^{0}_{B}$(cf.~\eq\eqref{eq:oB}),
we do not expect $\l$ to be ${\cal O}(10)$ and choose $\l=8$ to be the largest value of $\l$ used in the current computations.
}
Dashed curves in Fig.~\ref{fig:v2hydro} are corresponding to the upper and lower bound of the direct photon $v_2$ at $\pt =0.5$~GeV in the data\cite{Bannier:2014bja}(cf.~Fig.~\ref{fig:v2data}).
As one can see in Fig~\ref{fig:v2hydro},
depending on the value of $\o_B\tau_{\rel}$, 
the magnetic field would give contribution,
which is comparable to the data, to the soft photon $v_2$ for $\tau_B>2\sim 3$~fm.
We have also checked that for those $\tau_{B},\l$ which reproduce the photon $v_2$ in the experiment, 
our estimation on $\s_{0}$ based the photon production rate in the
absence of magnetic field will only be affected by $10-20\%$,
within the error bar shown in the Fig.~\ref{fig:pt05}.

\section{Summary and discussion}
\label{sec:summary}
In this paper,
we have estimated the electrical conductivity $\sigma$ in the quark gluon plasma(QGP) based on 
the soft photon production from the data and realistic hydrodynamic evolution.
We find that $\s/(e^{2}T)$ is in the range $0.4<\s/(e^{2}T)<1.1$.
Previously,
the electrical conductivity of QGP was mostly extracted from Euclidean correlator measured on lattice\cite{Ding:2010ga,Amato:2013naa,Brandt:2012jc,Aarts:2007wj}.
Those analyses always involve a non-trivial analytical continuation.
The present work offered an alternative estimation of the electrical conductivity.

Photon production in heavy-ion collisions have been studied extensively
(see for example Refs.~\cite{Chatterjee:2005de,*Dion:2011pp,*Shen:2013vja})
based on the thermal photon emission rate computed from perturbative QCD(pQCD)\cite{Arnold:2001ms}.
While one may apply pQCD at high photon energy,
its applicability for photon energy below a few GeV is not warranted. 
Indeed, 
experiment results\cite{Adare:2008ab} indicate that the hydrodynamic simulations with pQCD rate typically underestimate the thermal photon production.
In this work,
instead of taking the soft photon production rate of QGP as an input from certain microscopic calculations,
we have extracted such rate based on hydrodynamics and the data.  
It would be interesting to extend the method used in this paper to obtain information on photon production rate at other $\pt$ window. 

The effects of magnetic field on the photon production and photon azimuthal anisotropy, $v_2$, have attracted much attention recently
\cite{Tuchin:2010gx,Basar:2012bp,Tuchin:2012mf,Yee:2013qma}.
We hope the present study based on hydrodynamics 
would shed light on how sizable the effects of magnetic field would be.
In particular,
by computing the contributions from magnetic field to photon $v_2$ for various,$\tau_B$,
the life time of magnetic field in realistic hydrodynamic background,
we found that if the life-time of $\tau_B >2\sim 3$~fm,
the resulting soft photon $v_2$ is comparable to that measured in
experiment.
On the other hand,
if the life-time of magnetic field is as short as estimated in
Ref.~\cite{McLerran:2013hla},
the contribution from magnetic field to photon $v_2$ in low momentum
region might be negligible. 

It should be noticed that
magnetic field even in the absence of anomaly  would contribute to the photon $v_2$ via conventional synchrotron radiation\cite{Tuchin:2012mf}.
Distinguishing the effects of chiral anomaly is not that straightforward. 
In hydrodynamic regime,
however,
a model-independent conclusion can be drawn in the light of the \eq\eqref{eq:rateB}.
According to \eq\eqref{eq:rateB} and the discussion in Sec.~\ref{sec:v2B},
the contributions due to chiral anomaly are fully parametrized by the speed of chiral magnetic wave $\cmw$ while the effects due to the conventional cyclotron motion are parametrized by $\o_{B}\tau_{\rel}$.
Moreover,
the azimuthal angle dependence of the photon production is drastically affected by the additional pole structure of retarded Green's function due to the chiral magnetic wave.
For example, 
by Fourier transforming \eq\eqref{eq:rateB},
one can see explicitly that the Fourier component of $\cos(4\phi_{\vp})$ is proportional to $\cmw^2$,
suggesting that photon $v_4$ might be used to study the effects of chiral magnetic wave\cite{Yee:2013qma}.

We will conclude this paper by pointing out that chiral anomaly may play \textit{different} roles in soft photon production at RHIC and LHC.
At RHIC energy where  $\cmw$ is not very close to $1$, 
one may apply the approximate expression \eq\eqref{eq:rateB1}. 
At RHIC, 
we found that
suppression of the transverse conductivity due to Lorentz force may
play a dominant role to contribute to the photon  $v_2$.
However, 
at LHC energy where $\cmw$ approaches $1$ due to much larger magnetic field,
the pole of $G^{ij}_{R}(\o, \vp)$(cf.~\eq\eqref{eq:Formfactor}) corresponding to chiral magnetic wave will be very close to the light cone.
The photon production is largely enhanced along the direction $\hB$.
Physically,
this is due to the decay of the chiral magnetic wave into photon when $\cmw$ is close to $1$.
\footnote{We thank D. Kharzeev for pointing this out to us.}. 
This implies that chiral magnetic wave will give negative contribution to soft photon $v_2$
(see Ref.~\cite{Yee:2013qma} for a holographic example).
It is interesting to see if this will happen for soft photon production at LHC.

\acknowledgments
Y.Y. is indebted to Misha Stephanov for various discussion and suggestions during this project 
and Gocke Basar, Dimitri Kharzeev, Ho-Ung Yee for stimulating
discussion on photon production in the presence of magnetic field and
chiral anomaly.
Special thanks are devoted to Karl Landsteiner who pointed out a
mistake on the estimation of the longitudinal conductivity in the
previous version of this paper. 
Y.Y. would like to thank Chun Shen for e-mail correspondence on hydrodynamic simulations and Benjamin Bannier and
Richard Petti for conversations on photon production measured in experiment.
Y.Y. would like to express his gratitude to Jinfeng Liao, Lin Shu, Derek Teaney for helpful comments.
Y.Y. is in gratitude to the nuclear theory group of Stony Brook university for hospitality 
where this work was initiated and the funds from the Provost award of
UIC to support his visit to Stony Brook University. 
Y.Y. would also like to acknowledge 
the stimulating environment of the ``Quantum anomalies and hydrodynamics'' at Simons Center for
Geometry and Physics and of the``Thermal photons and dileptons in
heavy-ion collisions" workshop at RIKEN-BNL Research Center.
This work is supported by
the DOE grant No.\ DE-FG0201ER41195 and No.\ DE-AC02-98CH10886.

\bibliography{photon}
\end{document}